\documentstyle[aps,prl,epsfig]{revtex}

\begin{document} 
\wideabs{
\title{Vortex--vortex interaction in two-component Bose--Einstein condensates} 
\author{P. \"Ohberg$^{1,2}$ and L. Santos$^{1}$} 
\address{$^{1}$Institut f\"ur Theoretische Physik, Universit\"at Hannover, 
 D-30167 Hannover,Germany} 
\address{$^{2}$School of Physics and Astronomy, University of St Andrews, North Haugh,
St. Andrews, Fife KY16 9SS, Scotland} 
 
\maketitle 
 
\begin{abstract} 
The vortex--vortex interaction in two--component Bose--Einstein condensates is shown to
present characteristic effects not possible in single--component condensates. 
In particular, vortices in different components
undergo separate, but concentric, orbits for both same and 
opposite circulation. In addition, contrary to the well--known behavior in 
single--component condensates, vortices with the same circulation created in 
the less dense component merge in the presence of dissipation, 
allowing for the creation of metastable vortices with more than one quantum 
of circulation.

\end{abstract} 
} 


The experimental achievement of Bose-Einstein condensation (BEC) in trapped dilute alkali gases
 \cite{BEC} has opened the possibility to explore fundamental properties 
of low temperature physics in a very controllable way. 
In particular, superfluid phenomena, well known from $^4$He physics
\cite{Landau}, have been experimentally observed in BEC \cite{superfluidity}.
Among the phenomena related with superfluidity, the possibility of 
vortices with quantized circulation \cite{Donnelly}
has aroused a big interest.  Due to the diluteness of the atomic condensates, 
accurate theoretical predictions are possible.
In addition, in these systems, the 
diameter of the vortex core is typically three orders of magnitude larger than for 
superfluid Helium, allowing for a more straightforward optical observation.
Numerous theoretical schemes have been suggested for generating vortices in BEC, 
see e.g. Refs.\cite{Martzlin97,Dum98,Jackson98,Caradoc99,Dobrek99,Williams99,Feder99}.
So far only two techniques have successfully generated vortices
in condensates, namely the stirring  
with a blue detuned laser~\cite{Madison20}, and the coherent interconversion 
between the two components of a binary BEC~\cite{Matthews99}.
In addition the creation of a vortex ring after the decay of a dark soliton has 
been recently reported \cite{Anderson00}.
The dynamics and stability of vortices in trapped 
Bose--Einstein condensates has recently also been a subject of active investigation 
\cite{Fetterreview}.

In the recent years, the development of trapping techniques has allowed for the creation of  
multi-component condensates, which are formed by trapping atoms in different internal  
(electronic) states \cite{JILA2,MIT}. The multicomponent BEC,  
far from being a trivial extension of the single--component one, presents novel and fundamentally  
different scenarios for its ground state \cite{Ho,Patrik} and excitations \cite{Thomas}.  
In particular, it has been observed that the condensate can reach  
an equilibrium state characterized by the separation of the species in different domains \cite{MIT}. 
The realization of vortices in binary condensates in recent experiments at JILA 
\cite{Matthews99} has aroused the interest for the properties of vortices 
in multicomponent condensates \cite{Fetterreview,Ripoll99,McGee00}. Several 
interesting phenomena have been reported in this context, 
such as vortex ground states \cite{Ho96} and generation of persistent currents \cite{McGee00}.


In this Letter we discuss the interaction of vortices in a two--component BEC.
The first part of the Letter is devoted to homogeneous miscible binary condensates.
We consider in particular the case in which one vortex is created in each component.
In single--component BEC (or classical gas) in absence of dissipation,
two vortices with the same circulation orbit around each other, whereas 
vortices with opposite circulation move parallelly (vortex pair) \cite{Donnelly}. 
However, for the case of asymmetric nonlinear interactions between the two components
of a binary BEC, the velocity fields for the vortices are asymmetric. 
We show that in this scenario, vortices with the same circulation perform 
two concentric orbits. Interestingly, 
also vortices with opposite circulation perform concentric orbits, 
clearly differing from the one component behavior.

In the second part of this Letter we analyze the interaction of two vortices created in one of 
the components of an inhomogeneous binary BEC. 
In particular, we study the interaction of vortices with the same circulation, and show that 
the ``buoyancy'' effect \cite{McGee00} strongly distorts the vortex dynamics. 
In single--component BEC, vortices with the same circulation 
repel each other in the presence of dissipation, whereas
the opposite can be true in a two--component BEC.
This result strongly suggests that the creation of stable vortices with more than one quantum of 
circulation should be possible in a binary BEC.


Let us first consider the case of one vortex created in each of the components of 
a miscible binary BEC trapped in a 2D disc--shape box--like potential of radius $R$. 
The BEC can be considered homogeneous, except 
in a region of the size of the healing length, close to the box boundaries. 
The dynamics can be described in 2D if the mean--field interaction is smaller than the  
typical energy separation $E_{\perp}$ in the third direction \cite{Ketterle}.
To generate the vortices we use the phase--imprinting (PI) technique  
\cite{Dobrek99}, although other methods could be employed, as 
for example stirring methods \cite{Madison20}, or the method employed 
in Ref.\ \cite{Matthews99}, which 
already involves two components.  

For sufficiently low temperatures the dynamics is well  
described by two coupled Gross--Pitaevskii equations 
\begin{eqnarray} 
i\hbar\frac{\partial}{\partial t}\psi_j({\bf r},t)&=& \left  
\{ -\frac{\hbar^2}{2m}\nabla^2+V({\bf r})+  
g_{jj}|\psi_j({\bf r},t)|^2+ \right. \nonumber \\ 
&&\left. g_{jl}|\psi_l({\bf r},t)|^2  
\right \} 
\psi_j({\bf r},t), 
\label{GPE} 
\end{eqnarray} 
where $V({\bf r})$ is the trap potential, $g_{jl}=4\pi\hbar^2a_{jl}/mL$ is  
the coupling constant 
(averaged over the frozen--out transversal direction)
between the components $j$ and $l$ with the transversal length $L$,  
$m$ is the atomic mass and $a_{jl}$ the scattering length between $j$ and $l$ ($j,l=1,2$).


We have numerically simulated the creation of one vortex in each component 
by using PI. The creation of a vortex in one component modifies the density in the other  
one, due to the coupling in Eq.\ (\ref{GPE}). For example, dark solitons in homogeneous two-component BEC 
can only be created at distances comparable to the healing length \cite{Letter}, 
$l_0=\hbar/\sqrt{gnm}$, where $n$ is the density of the sample. The latter does not apply to
vortices, which can be created at distances significantly larger than $l_0$, without 
being destroyed by the density fluctuations of the other component.
Although we have numerically observed stable creation of vortices at distances larger than $20l_0$, typical 
distances of the order of $8l_0$ are used in the simulations below, 
since the time scale of the vortex--vortex dynamics is given by $md^2/\hbar$, where $d$ is the 
distance between the vortices. Therefore shorter distances allow for faster dynamics. This facilitates 
the numerical simulations, and reduces instabilities during the vortex interaction.
As already observed in recent experiments \cite{Matthews99}, the vortex core in one component 
is filled by the other component. Although the two components interact 
through the density, and therefore they are not directly experiencing the circulation of the other vortex,
the filling of each core does feel the circulation created by the vortex in the same component, 
dragging in its way the vortex core in which it is placed. 
The result is an indirect interaction.

Let us postpone the discussion about the boundary effects and concentrate on the 
situation in which the radius $R$ of the disc--shape trap
is sufficiently large to neglect the  
boundary effects during the time scale of the simulations. 
In the following we consider $g_{11}=g_{22}=1.05 g_{12}$. 
These particular $g_{ij}$ values have been chosen 
for simplicity, to guarantee a complete overlap between the wavefunctions of both components.
In the case of $^{87}$Rb,  the coupling constants have the values 
$g_{11}:g_{22}:g_{12}\equiv 0.97:1.03:1.0$. These values also imply a 
large overlap, without separate domains, and therefore the dynamics we discuss below
should be observable in $^{87}$Rb.

In the symmetric situation, $n_{1,2}=n_0$, the qualitative behavior of single--component 
vortex--vortex interaction is recovered, namely, vortices with the same circulation 
rotate around each other, forming a vortex orbit, whereas vortices of opposite circulation 
move paralelly (vortex pair) through the fluid. The time evolution of the vortex cores 
is, however, different to the single--component one, since  
the effective masses of the vortex cores are different.

\begin{figure}[ht] 
\begin{center}\ 
\epsfxsize=4.5cm 
\hspace{0mm} 
\psfig{file=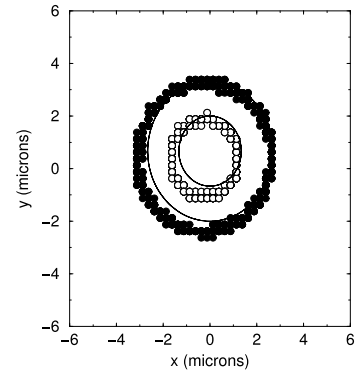,width=4.5cm}\\[0.1cm] 
\end{center} 
\caption{Numerical (circles) and analytical (solid lines) results 
for the vortex orbits with $n_1=2n_2$, $d=4\mu$m, $R=20$d. 
The denser component occupies the inner orbit.}
\label{fig:1}  
\end{figure}

This picture breaks down when the $n_1\neq n_2$.
Fig.\ \ref{fig:1} shows the position of the vortex lines for 
the case of $n_1=2n_2=10^{11} cm^{-2} $, $d=4 \mu$m, and $R=20d$, 
for vortices with the same circulation. In this asymmetric scenario,  
contrary to the case of single--component vortices, 
vortices with the same circulation do not undergo an orbit, but two concentric ones.
The inner orbit corresponds to the vortex in the denser component, whereas the 
outer one corresponds to the vortex created in the less dense one.
Interestingly, such a behavior is also present for the case of vortices with opposite circulation.


In order to understand the formation of the planetary orbits when $n_1\not= n_2$, let us 
consider the situation of a vortex line in the component $1$ ($2$) at a  
position $\vec r_1$ ($\vec r_2$). The density of the component $1$ ($2$) is considered constant 
and equal to $n_1$ ($n_2$) everywhere in the system, except in the core of the vortices, which 
will be modeled as a cylinder of radius $l_1$ ($l_2$). In the core of the vortex in the component 
$1$ ($2$) the density of $1$ ($2$) is considered zero, 
whereas the density of $2$ ($1$) is considered $n_1+n_2$, in such a way that the total density is flat. 
In the following calculation, we shall assume that $|\vec r_1|,|\vec r_2| \ll R$, 
i.e. the vortices are placed close to the trap center,  and the vortex cores are far away from 
each other, $l_1,l_2 \ll |\vec r_2-\vec r_1|=|\vec r|$.
The conservation of linear momentum and energy in the system, up to order ${\cal O}((r_j/R)^2)$, 
${\cal O}((l_j/r)^2)$, ($j=1,2$), lead respectively to the following equations:
\begin{eqnarray}
&& n_1 \vec u_1 + \kappa n_2 \vec u_2 = 
\frac{n_2 l_2^2+\kappa n_1 l_1^2}{r^2}
\left (\vec u -2\left (\frac{\vec r \cdot \vec u}{r^2}\right )\vec r \right ), \label{momcon}\\
&& \left (\frac{r}{R} \right )^2 (n_1\vec r_1\cdot\vec u_1+n_2\vec r_2\cdot\vec u_2)
=-\frac{n_2 l_2^2+n_1 l_1^2}{r^2} (\vec r\cdot\vec u), \label{encon}
\end{eqnarray}
where the velocity of the vortex in the $j$--th component is given by 
$\vec v_j=\vec u_j / r^2$, $\vec u=\vec u_2-\vec u_1$, and $\kappa=1$ ($-1$) for vortices with the 
same (opposite) circulation. Neglecting in Eq.\ (\ref{momcon}) the relatively small rhs, 
and assuming $R$ tending to $\infty$, one obtains the solution 
$\vec v_2 \sim (\vec r_1-\vec r_2)^{\perp}/|\vec r_1-\vec r_2|^2$, and 
$\vec v_1 = -(\kappa n_2/n_1)\vec v_2 $ \cite{footnote}.
The motion of the vortex lines is then determined by the first--order equations 
$d\vec r_j/dt=\vec v_j$. The trajectories provided by these equations 
(solid lines in Fig.\ \ref{fig:1}) describe very well the numerically calculated vortex orbits.
Note that since the velocity field is always tangential to the vector which joins the 
vortex lines, the distance between the vortices remains constant at any time, 
and consequently the angular frequency for both orbits is the same.
Therefore, vortices of the same circulation are always at opposite angles  
of their respective orbits, whereas vortices with opposite circulation always occupy the 
same angle. Hence, the angular frequency for both orbits is the same. 
From this fact, it is easy to show that for vortices with the same/opposite 
circulation, initially placed at a distance $d$, the orbit of the outer vortex 
(say in component $1$) has a radius $r_1=d n_2/(n_1\pm n_2)$, whereas the
inner one has a radius $r_2=d n_1/(n_1\pm n_2)$. Note that for $n_1=n_2$ the familiar one--component 
results are retrieved.

\begin{figure}[ht] 
\begin{center}\ 
\epsfxsize=4.5cm 
\hspace{0mm} 
\psfig{file=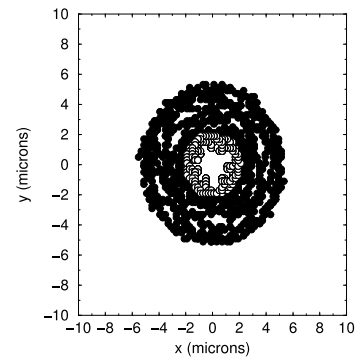,width=4.5cm}\\[0.1cm] 
\end{center} 
\caption{Same as Fig.\ \ref{fig:1} but with $R=4.5d$.}
\label{fig:2}  
\end{figure}

Let us discuss at this point the role of the boundaries on the dynamics.
Our numerical analysis shows that 
the finite size of the trap does not disturb the distance between the vortices, 
and, as in the case of $R\rightarrow\infty$,  $\vec r_1 -\vec r_2$ moves in a circle. 
However, the coordinate $(\vec r_1 +\vec r_2)/2$, which for $R\rightarrow\infty$ moves in a 
circle as well, performs in the presence of a finite $R$ a cycloid motion. 
As a result, the planetary orbits are not circular anymore, but will precess around a 
circular orbit, as one can observe 
in Fig.\ \ref{fig:2}, where the same parameters of Fig.\ \ref{fig:1} are considered, except 
$R=4.5d$. Nevertheless, similar to the case $R\rightarrow\infty$,
the trajectory of the vortex in the denser component is always the inner trajectory.


In the last part of this Letter we analyze the case of 
a binary BEC trapped in a 2D harmonic potential 
of frequency $\omega$.  This resembles the situation in which vortices
are currently created in $^{87}$Rb in JILA \cite{Matthews99}.
In particular, we shall analyze the interaction of two vortices 
in the same component, which are created 
symmetrically from the condensate center, and with the same circulation.
In homogeneous as well as single--component trapped condensates, and in the 
presence of dissipation, the vortices will spiral out from their initial 
orbit until they finally reach the boundaries of the condensate cloud, where they are 
destroyed \cite{Fedichev}. 
We shall show that this is not the case in an inhomogeneous binary BEC.

\begin{figure}[ht] 
\begin{center}\ 
\epsfxsize=5.0cm 
\hspace{0mm} 
\psfig{file=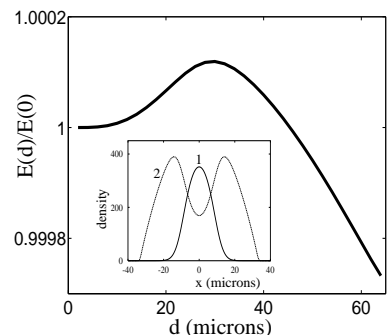,width=5.0cm}\\[0.1cm] 
\end{center} 
\caption{Energy as a function of vortex distance (see text for details). 
Inset: Density distribution without vortices.}
\label{fig:3}  
\end{figure}

In a recent paper \cite{McGee00}, the dynamics of a filled vortex in a trapped binary BEC was 
analyzed. 
In an inhomogeneous potential, the lowest energy solution is reached when
the component with smaller nonlinear self--interaction occupies the regions of 
highest density, i.e. the trap center.
Therefore, in our case, if the less dense component is displaced from the trap center, 
it will tend to return to it. This effect is sometimes called the 
``buoyancy'' effect \cite{McGee00}. 
This effect can counteract the tendency of an off--centered vortex 
to spiral out. In particular, it has been shown that 
for sufficiently large nonlinearity a vortex, created in the component with larger 
self--interaction at the center of the trap, can become 
metastable \cite{McGee00}. We employ a similar treatment to analyze 
the interaction of vortices. It consists in evaluating, for different 
distances between vortices, the free energy  
\begin{eqnarray}
E&=&\int d^3{\bf r}\,
\Bigl(\frac{\hbar^2}{2m}\bigl(|\nabla\psi_1({\bf r})|^2
+|\nabla\psi_2({\bf r})|^2\bigr) \nonumber\\ &&+
V_1({\bf r})|\psi_1({\bf r})|^2 +
V_2({\bf r})|\psi_2({\bf r})|^2 \nonumber\\ && +
\frac{g_{11}}{2}|\psi_1({\bf r})|^4
+g_{12}|\psi_1({\bf r})|^2|\psi_2({\bf r})|^2
+\frac{g_{22}}{2}|\psi_2({\bf r})|^4\Bigr).
\label{energy}
\end{eqnarray}
In order to obtain a clear value of the free energy associated with every 
vortex configuration, instead of creating the vortices using the PI method discussed
above, we generate them, as in Ref.\ \cite{McGee00}, by evolving the 
Gross--Pitaevskii equation (\ref{GPE}) in imaginary time, fixing at every iteration 
the phase of the component in which the vortices are created, 
in order to have the correct phase properties.
Once imprinted in this way two vortices at position $(-d/2,0)$ and $(d/2,0)$, 
we calculate the energy $E(d)$ following Eq.\ (\ref{energy}).
For the case of single component condensates, one can show that $E(d)$ decreases with $d$, for any
$d$. This reflects the fact that in the presence of dissipation the vortices tend 
to spiral out. For vortices created in the denser component (say $1$), 
the situation is, however, completely different. Fig.\ \ref{fig:3} shows the 
case of $^{87}$Rb ($a_{11}=1.03a_{12}$, $a_{22}=0.97a_{12}$, and 
$a_{12}=5.7$nm), with a trap frequency $\omega=2 \pi\times 30$Hz
and $L=1\mu$m, with number of particles $N_1=10^5$ and $N_2=8\times 10^5$. 
One observes that $E(d)$ clearly 
grows with $d$ for low $d$, until it reaches a maximal value. This maximum 
corresponds to the case in which the vortices are placed at the maxima 
of the unperturbed density distribution of the component $1$ (see inset in Fig.\ \ref{fig:3}). 
For larger distances the tendency of the vortices to spiral 
out is no more counteracted since the slope of the density distribution 
changes the sign. Therefore, vortices created sufficiently close 
to the center will tend to spiral inwards in the presence of dissipation, until 
eventually merging at the trap center. Once they merge, a vortex with double circulation 
will be created. Since due to its topology a vortex can only be destroyed if it
reaches a region of zero density, the created vortex of double 
circulation will remain metastable at the trap center, where only inelastic 
processes will in practice limit its lifetime. Following similar arguments, 
the merging of $n_v$ vortices should allow for the creation of a metastable 
vortex with $n_v$ quanta of circulation, opening a new scenario which is naturally 
forbidden in both homogeneous single-- and multi--component BEC, as well as 
in inhomogeneous single--component condensates.

Let us finally point out, that a similar argument applies for vortices with opposite circulation 
created in the denser component and moving parallelly in a toroidal trap \cite{Arlt}. 
In that case, the ``buoyancy'' effect counteracts the natural tendency of the vortices to  
mutually annihilate, allowing for the creation of a vortex pair,
which will circulate around the torus center in a metastable orbit.


Summarizing, we have shown that the vortex--vortex interaction in multicomponent condensates, 
presents characteristic features such as planetary orbits for 
vortices placed in different components, and opens new interesting possibilities 
for creating metastable vortices with more than one quanta of circulation, as well as 
metastable vortex pairs in toroidal geometries.

The analysis of the vortex--vortex interaction has been limited 
to the case of 2D traps.
Lower dimensional condensates are currently actively investigated, and very recently 
the experimental observation of 2D (and even 1D) condensates have been reported \cite{Simo,Ketterle}.
However, the results of the present Letter, open interesting questions
concerning the behavior of vortices in 3D binary BEC.
In particular, for linear vortices,  
if the relative density between the different components 
varies in the direction of the vortex lines, it should be expected that vortices 
created in different components will undergo at each transversal position  
planetary orbits of different periods and radii. Consequently, non trivial 
distortions of the vortex lines are expected.
In addition, the recent observation of vortex--rings \cite{Anderson00} in condensates
stimulates the investigation of the interaction 
of such more complicated structures in binary BEC.
The analysis of these effects is beyond the scope of the present Letter, and will be 
the subject of further investigation.

We acknowledge  support from Deutsche Forschungsgemeinschaft (SFB 407),  
TMR ERBXTCT-96-002, ESF PESC BEC2000+ and EPSRC. 
L. S. wishes to thank the Alexander von Humboldt Foundation, the Federal Ministry of Education and 
Research and the ZIP Programme of the German Government.
Discussions with T. Busch, M. Lewenstein and Anna Sanpera are acknowledged.

\begin{references}
\bibitem{footnote} The proportionality constant in $\vec v_2$ determines the time scale of the 
vortex dynamics, but it is not important for the determination of the vortex orbits.
\end{references}
\end{document}